\documentclass[twocolumn,showpacs,preprintnumbers,amsmath,amssymb,superscriptaddress]{revtex4}

\usepackage{graphicx}
\usepackage{dcolumn}
\usepackage{bm}

\begin{document}

\title{Negative differential conductance and magnetoresistance oscillations due to
spin accumulation in ferromagnetic double-island devices}

\author{Ireneusz Weymann} \email{weymann@amu.edu.pl}
\affiliation{Department of Physics, Adam Mickiewicz University,
61-614 Pozna\'n, Poland}
\author{J\'ozef Barna\'s}
\affiliation{Department of Physics, Adam Mickiewicz University,
61-614 Pozna\'n, Poland} \affiliation{Institute of Molecular
Physics, Polish Academy of Sciences, 60-179 Pozna\'n, Poland}

\date{\today}

\begin{abstract}
Spin-dependent electronic transport in magnetic double-island
devices is considered theoretically in the sequential tunneling
regime. Electric current and tunnel magnetoresistance are analyzed
as a function of the bias voltage and spin relaxation time in the
islands.  It is shown that the interplay of spin accumulation on
the islands and charging effects leads to periodic modification of
the differential conductance and tunnel magnetoresistance. For a
sufficiently long spin relaxation time, the modulations are
associated with periodic oscillations of the sign of both the
tunnel magnetoresistance and differential conductance.
\end{abstract}


\maketitle

{\it Introduction} -- Spin-dependent transport in single-electron
devices is attracting currently much attention from both
fundamental and application points of view
\cite{ono96,barnas98,takahashi98,imamura99,shimada02,weymann03,ernult04,yakushiji05}.
It has been shown that the interplay of single-electron charging
effects and spin dependence of tunneling processes (caused by
ferromagnetism of external and/or central electrodes) in
single-electron transistors gives rise to novel interesting
phenomena, such as periodic modulation of the tunnel
magnetoresistance (TMR) with increasing either bias or gate
voltages, quasi-oscillatory behavior of the spin accumulation on
the central electrode (referred to as an island in the following),
enhancement of TMR in the Coulomb blockade regime, and others.
Some of the theoretically predicted phenomena have already been
confirmed experimentally.

Recently, transport characteristics of granular magnetic
nanobridges connecting two external magnetic and/or nonmagnetic
electrodes were investigated experimentally
\cite{imamura00,takanashi00,yakushiji02}. The nanobridges
consisted of magnetic grains distributed in a nonconducting
matrix. The corresponding current-voltage (I-V) curves revealed
characteristic steps due to discrete charging of the grains with
single electrons (Coulomb steps). Apart from this, oscillations of
the TMR effect (associated with the transition from one magnetic
configuration to the other) with the bias voltage have also been
observed \cite{yakushiji02}. Moreover, two additional features of
the transport characteristics were found, whose physical mechanism
and origin needs further considerations. First, the differential
conductance was found to change sign periodically with increasing
bias voltage \cite{imamura00,takanashi00}. Second, the TMR was
shown to oscillate between positive and negative values with
increasing bias voltage. Physical origin of these oscillations,
however, remained unclear, although the role of spin accumulation
was invoked following the results obtained for a single-island
system \cite{barnas98}.

For small lateral and vertical dimensions of the nanobridges,
electronic transport between the external leads occurs as a result
of consecutive electron jumps {\it via} metallic grains, and the
corresponding electronic paths can include either one, two, or
more grains. Transport through a single grain and the associated
TMR have already been analyzed \cite{barnas98}, but the results
cannot describe the above mentioned data obtained on magnetic
multi-grain nanobridges. However, the experimental setup described
above can be modelled by assuming only two grains involved in the
transport processes \cite{imamura00}. Although the Coulomb steps
have been accounted for by the existing theoretical models based
on discrete charging with single electrons, the origin of negative
differential conductance (NDC) as well as the oscillations in TMR
with alternating sign remains unclear. Therefore, this problem is
considered in this paper in more details.

\begin{figure}[t]
\includegraphics[width=0.66\columnwidth]{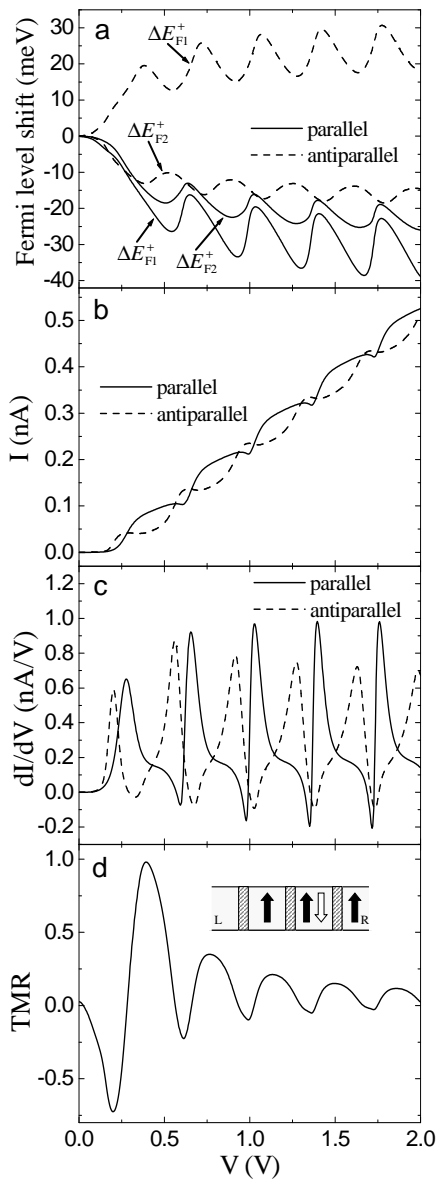}
  \caption{\label{Fig:characteristics} Shifts of the Fermi levels
  for spin-majority electrons (a) and currents (b)
  in the P and AP configurations;
  differential conductance in the AP configuration (c);
  and TMR (d) as a function of the bias voltage  for
  $\tau_{\rm sf,1}=\tau_{\rm sf,2}\rightarrow\infty$.
  The other parameters are: $T=140$ K, $C_{\rm L}=0.45$ aF,
  $C_{\rm M}=0.2$ aF, $C_{\rm R}=0.35$ aF,
  $\beta_1=\beta_2=\beta_{\rm R}=0.2$, and $\beta_{\rm L}=1$. The total barrier resistances
  in the P configuration are: $R_{\rm L}/3500=
  R_{\rm M}=R_{\rm R}=1$ M$\Omega$.
  In the AP configuration, $R_{r,\uparrow}^{\rm AP}=R_{r,\downarrow}^{\rm AP}
  = (R_{r,\uparrow}^{\rm P}R_{r,\downarrow}^{\rm P})^{1/2}$, for $r={\rm M,R}$.
  The parameters correspond to the experimental ones taken from Ref.~\onlinecite{imamura00}.}
\end{figure}

Following the above discussion, we consider transport through a
double-island device consisting of two metallic grains attached to
two external leads. The two islands are separated from each other
and also from the external leads by tunnel barriers. In a general
case all the four metallic components of the device (two external
electrodes and two islands) can be ferromagnetic
\cite{ono96,shimada02}, but in the following we will consider the
situation with one external lead being nonmagnetic and the other
three components being ferromagnetic. We also restrict our
considerations to collinear [parallel (P) and antiparallel (AP)]
alignments of the magnetic moments, as indicated in the inset of
Fig.~\ref{Fig:characteristics}(d).

The key role in our analysis is played by the nonequilibrium spin
accumulation, which occurs for a sufficiently long spin relaxation
time on the islands. Strictly speaking, the accumulation may take
place when the spin relaxation time is significantly longer than
the time between successive tunneling processes. Assuming this is
the case, we show that the interplay of charging effects and spin
accumulation gives rise to both effects described above, i.e., to
the NDC and periodic change of the sign of TMR with increasing
bias voltage. This behavior of transport characteristics  accounts
for the experimental observations. We also show that the presence
of NDC depends on the magnetic configuration of the device and by
switching from one configuration to the other one may change sign
of the differential conductance (eg., by an external magnetic
field). This behavior may be of some interest for possible future
applications in spintronics/magnetoelectronics devices.

{\it Model} -- The system adapted to model the phenomena discussed
above consists of the left and right leads and two central
islands, separated from each other and from the leads by tunnel
barriers, as shown schematically in the inset of
Fig.~\ref{Fig:characteristics}(d). The bias voltage is applied to
the external leads -- $V_{\rm L}$ to the left one and $V_{\rm R}$
to the right one. In a general case one can attach capacitively
gate voltages to both islands. However, to model the experimental
results described above, we neglect the gate voltages in the
following. We assume that the size of each island is sufficiently
small to have the charging energies of the islands significantly
larger than the thermal energy $k_{\rm B}T$, but still large
enough to neglect size quantization. Apart from this, we assume
that the barrier resistances are much larger than the quantum
resistance, $R_{r}\gg h/e^2$ ($r={\rm L,M,R}$). The system is then
in a well defined charge state described by $n_1$ and $n_2$ excess
electrons on the first (left) and second (right) islands,
respectively, and the electrostatic energy of the system is given
by the formula \cite{wiel,weymann04}
\begin{eqnarray}\label{Eq:Ech}
  E(n_{1},n_{2})&=&E_{\rm C_1}\left( n_{1} - \frac{C_{\rm L}V_{\rm L}}{e}
  \right)^2+ E_{\rm C_2} \left( n_{2} - \frac{C_{\rm R}V_{\rm R}}{e} \right)^2 \nonumber\\
  &&+2E_{\rm C_M}\left( n_{1} - \frac{C_{\rm L}V_{\rm L}}{e} \right) \left(
  n_{2} -\frac{C_{\rm R}V_{\rm R}}{e} \right)\,,
\end{eqnarray}
where $C_{\rm L(R)}$ is the capacitance of the left (right)
junction, $E_{\rm C_1}$ and $E_{\rm C_2}$ denote the charging
energies of the two islands, $ E_{\rm C_1(C_2)}=e^2/(2C_{1(2)}) [1
- C_{\rm M}^2/ (C_1 C_2)]^{-1}$, and $E_{\rm C_M}$ is the energy
of electrostatic coupling between the islands, $E_{\rm
C_M}=e^2/(2C_{\rm M})(C_1 C_2 /C_{\rm M}^2-1 )^{-1}$, with
$C_{1(2)}$ being the total capacitance of the first (second)
island, $C_{1(2)}=C_{\rm L(R)}+C_{\rm M}$, and $C_{\rm M}$
denoting the capacitance of the middle junction (the one between
the islands). This allows us to employ the quasi-classical theory
based on the master equation and the Fermi golden rule for the
tunneling rates \cite{grabert}. Such approach describes well
transport in the sequential tunneling regime and corresponds to
taking into account only the first-order tunneling processes,
which are exponentially suppressed in the Coulomb blockade regime
but give the dominant contribution to electric current when the
applied voltage exceeds a certain threshold. Moreover, we take
into account only non-spin-flip (spin-conserving) tunneling
processes through the barriers.

When a bias voltage $V=V_{\rm L}-V_{\rm R}$ is applied to the
system and spin relaxation time on the islands is sufficiently
long, a nonequilibrium magnetic moment can accumulate on each
island. Let us denote the corresponding shift of the Fermi level
for electrons with spin-$\sigma$ on the $j$-th island by $\Delta
E_{{\rm F}j}^\sigma$. When in the initial state there were $n_1$
and $n_2$ excess electrons on the islands, the spin-dependent
tunneling rate from the left electrode to the first island is then
given by the formula
\begin{eqnarray}
\Gamma_{\rm L1}^\sigma(n_1,n_2)= \frac{1}{e^{2}R_{\rm
L}^\sigma}\frac{\Delta E_{\rm L1}^\sigma(n_1,n_2)}{\exp \left[
\Delta E_{\rm L1}^\sigma(n_1,n_2)/k_{\rm B}T\right] -1}\;,
\end{eqnarray}
where $R_{\rm L}^\sigma$ is the spin-dependent resistance of the
left barrier and $\Delta E_{\rm L1}^\sigma(n_1,n_2)$ describes the
change in the electrostatic energy of the system caused by the
respective tunneling event, $\Delta E^\sigma_{{\rm
L1}}(n_{1},n_{2})= E(n_{1}+1,n_{2})-E(n_{1},n_{2})+eV_{\rm
L}+\Delta E_{\rm F1}^\sigma$. (According to our notation the
electron charge is $-e$ with $e>0$.)

For the following discussion it is convenient to distinguish
between spin projection on the global quantization axis ($\sigma
=\uparrow ,\downarrow$) and on the local quantization axis
($\sigma =+,-$, with $\sigma =+$ and $\sigma =-$ corresponding to
spin-majority and spin-minority electrons, respectively). The spin
asymmetry in tunneling processes follows from spin-dependent
density of states and spin-dependent tunneling matrix elements.
Let us  define the parameters $\beta_r=D_r^+/D_r^-$ for the leads
($r ={\rm L,R}$) and $\beta_j=D_{{\rm I}j}^+/D_{{\rm I}j}^-$ for
the islands ($j=1,2$), where $D_{r}^{+(-)}$ and $D_{{\rm
I}j}^{+(-)}$ are the appropriate densities of states for
spin-majority (spin-minority) electrons in the leads ($r={\rm
L,R}$) and islands ($j=1,2$), respectively. Thus, the Fermi level
shifts due to spin accumulation on the islands obey the condition
$\Delta E^+_{{\rm F}j}/\Delta E^-_{{\rm F}j} =-\beta_j$ for the
first ($j=1$) and second ($j=2$) islands.

The spin asymmetry of the barrier resistances can be described by
the parameters $\alpha_r=R_{r}^{\uparrow}/R_{r}^{\downarrow}$ for
$r={\rm L,M,R}$. Assuming a constant (independent of energy)
density of states and constant (independent of energy and spin)
matrix elements, one can write $\alpha_r=D_{r}^{\downarrow}
D_{{\rm I} j}^\downarrow/D_{r}^{\uparrow}D_{{\rm I} j}^\uparrow$
for the left ($r={\rm L}$, $j=1$) and right ($r={\rm R}$, $j=2$)
barriers, and $\alpha_r=D_{{\rm I} 1}^\downarrow
D^\downarrow_{{\rm I} 2}/D^\uparrow_{{\rm I} 1}D_{{\rm I}
2}^\uparrow$ for the central ($r=M$) barrier. When the local and
global spin quantization axes coincide (parallel configuration),
then one can write $\alpha_r=1/(\beta_r\beta_j)$ for the left
($r={\rm L}$, $j=1$) and right ($r={\rm R}$, $j=2$) barriers, and
$\alpha_r=1/(\beta_1\beta_2)$ for the central ($r={\rm M}$)
barrier. The above formulas are also applicable to the situation
with magnetic moment of a lead or an island reversed (antiparallel
configuration), but with the corresponding $\beta$ replaced by
$1/\beta$.

The probability $P(n_1,n_2)$ that the system is in a charge state
$(n_1,n_2)$ can be determined in a recursive way from the
appropriate steady-state master equation \cite{weymann04}. The
electric current flowing through the left junction is then given
by
\begin{eqnarray}
  I_{\rm L}= -e\sum\limits_{\sigma}\sum\limits_{n_{1},n_{2}}
  \left[\Gamma_{\rm L 1}^\sigma(n_{1},n_{2})-\Gamma _{\rm 1
  L}^\sigma(n_{1},n_{2})\right] P(n_{1},n_{2})\,.
\end{eqnarray}
The associated shifts of the Fermi level can be calculated in a
self-consistent way from the relations \cite{barnas99}
\begin{equation}
  \frac{1}{e}\left( I_{\rm M(R)}^\sigma - I_{\rm L(M)}^\sigma
  \right) - \frac{D_{{\rm I}1(2)}^{\sigma}\Omega _{\rm
  I1(2)}}{\tau_{\rm sf,1(2)}} \Delta E_{\rm F1(2)}^{\sigma}=0 \,,
\end{equation}
where $\Omega_{{\rm I}j}$ is the volume of the island $j$,
$\tau_{{\rm sf},j }$ denotes the spin-flip relaxation time in the
$j$-th island, and $I_r^\sigma$ is the current flowing through the
barrier $r$ ($r={\rm L,M,R}$) in the spin channel $\sigma$.

{\it Results and discussion} -- In Figure
\ref{Fig:characteristics} we show transport characteristics of a
device whose two islands as well as the right electrode are made
of the same ferromagnetic metal, whereas the left electrode is
nonmagnetic, as shown in the inset of
Fig.~\ref{Fig:characteristics}(d), where also both parallel and
antiparallel configurations are defined. This system geometry
corresponds to the situation studied experimentally in
Ref.~[\onlinecite{imamura00}]. The transport characteristics have
been obtained for the limit of long spin relaxation time. The
current-induced shifts of the Fermi level (due to spin
accumulation) for the spin-majority electrons in both islands are
shown in Fig.~\ref{Fig:characteristics}(a) for the P and AP
configurations and for positive bias voltage (electrons flow from
right to left). A nonzero spin accumulation occurs in both
magnetic configurations. In the P configuration the shift of the
Fermi level for spin-majority electrons is negative for both
islands, whereas in the antiparallel configuration it is positive
for the first island and negative for the second one. Such a
behavior can be accounted for by taking into account spin
asymmetries of the tunneling processes through all the three
barriers, similarly as it was done in the case of a double-barrier
system \cite{barnas98}.

The Coulomb steps in the P and AP configurations are significantly
different [see Fig.~\ref{Fig:characteristics}(b)]. There are two
reasons of this difference. First, the overall resistance of a
given barrier depends on the relative orientation of the magnetic
moments of adjacent ferromagnetic components of the device.
Second, the spin accumulations in the P and AP configurations are
also different. The latter fact is of particular importance for
the present analysis.

The differential conductance corresponding to the $I-V$ curves
from Fig.~\ref{Fig:characteristics}(b) is shown in
Fig.~\ref{Fig:characteristics}(c). In both P and AP configurations
the differential conductance changes sign periodically with
increasing bias. However, the bias voltage range of NDC for the P
configuration is different from that for the AP one. The
corresponding phase difference varies with the bias voltage.
Furthermore, NDC is more pronounced in the AP configuration and
its absolute magnitude increases with increasing bias voltage, as
shown in Fig.~\ref{Fig:characteristics}(c). This behavior of NDC
is consistent with experimental observation reported in
Ref.~\onlinecite{imamura00}.

\begin{figure}[b]
\includegraphics[width=0.7\columnwidth]{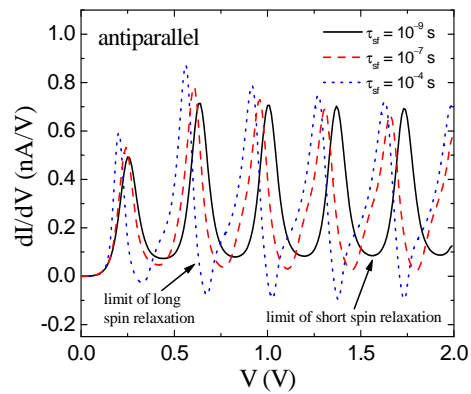}
  \caption{\label{Fig:G_AP} (color online) Differential conductance in
  the antiparallel configuration as a function
  of the bias voltage calculated for different values of the spin relaxation time
  $\tau_{\rm sf,1}={\tau_{\rm sf,2}}=\tau_{\rm sf}$
  and for $D_{\rm I1}^+ \Omega_{\rm I1}=D_{\rm I2}^+ \Omega_{\rm I2}=1000$/eV.
  The other parameters are the same as in Fig.~\ref{Fig:characteristics}.}
\end{figure}

The TMR  effect associated with the transition from antiparallel
to parallel configurations is visualized in
Fig.~\ref{Fig:characteristics}(d). The TMR is periodically
modulated with increasing bias voltage. The initial phase of the
oscillations depends on the system parameters and can change from
negative to positive. Moreover, the modulations are associated
with a periodic change of TMR between positive and negative
values.

Oscillations of the sign of TMR and differential conductance
result from spin accumulations in both islands and are absent in
the limit of fast spin relaxation (no spin accumulation) on the
islands. The results presented in Fig.~\ref{Fig:characteristics}
have been calculated for the long spin relaxation limit. In such a
limit some spin accumulation may occur even for a very small
current flowing through the system, giving rise to  NDC and
oscillations of the TMR sign for small bias voltages. However,
both effects disappear when the spin relaxation time is shorter
than the time between successive tunneling events. Thus, for a
finite relaxation time, one may expect no NDC and no TMR
oscillations for small voltages and an onset of these effects at
larger voltages. This is because at some voltage there is a
crossover from the fast to slow spin relaxation limits. In fact
such a behavior is consistent with experimental data presented in
Ref.~\onlinecite{imamura00}.

\begin{figure}[t]
\includegraphics[width=0.7\columnwidth]{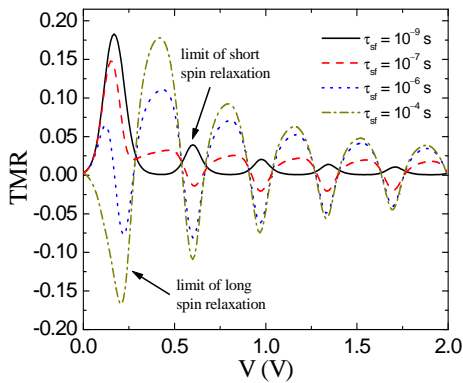}
  \caption{\label{Fig:TMR} (color online) The TMR effect as a function of
  the bias voltage calculated for different values of spin relaxation time
  $\tau_{\rm sf,1}={\tau_{\rm sf,2}}=\tau_{\rm sf}$
  and for $D_{\rm I1}^+ \Omega_{\rm I1}=D_{\rm I2}^+ \Omega_{\rm I2}=1000$/eV.
  The other parameters are the same as in Fig.~\ref{Fig:characteristics}.}
\end{figure}

The disappearance of NDC with decreasing spin relaxation time
$\tau_{\rm sf}$ is shown explicitly in Fig.~\ref{Fig:G_AP}, where
the bias dependence of differential conductance is presented for
different values of the spin relaxation time. This figure clearly
shows that NDC disappears when spin relaxation time decreases, in
agreement with the  above discussion. In the limit of fast spin
relaxation time, the differential conductance is positive,
although its periodic modulation still remains.

Similarly, periodic oscillations of the sign of TMR also disappear
with decreasing spin relaxation time $\tau_{\rm sf}$. This
behavior is shown in Fig.~\ref{Fig:TMR}, where the bias dependence
of TMR is shown for several values of $\tau_{\rm sf}$. First, the
transitions to negative TMR disappear with decreasing $\tau_{\rm
sf}$. The TMR becomes then positive, although some periodic
modulations survive. Second, the phase of the modulations shifts
by about $\pi$ when the spin relaxation varies from fast to slow
limits.

In conclusion, we have analyzed transport through double-island
device and shown that the negative differential conductance
measured experimentally is due to nonequilibrium spin accumulation
in the islands. Furthermore, spin accumulation may also lead to
oscillations in TMR between negative and positive values. The
effect of NDC occurs in both configurations and transition from
one configuration to the other may result in transition from
positive to negative differential conductance, which may be of
some importance for applications in spintronics devices.

The work was supported by the Polish State Committee for
Scientific Research through the projects PBZ/KBN/044/P03/2001 and
2 P03B 116 25, and by EU through RTNNANO contract No
MRTN-CT-2003-504574.


\end{document}